\documentstyle[12pt,graphicx]{article}
\begin{document}
\title{The Euclidean Scalar Green Function in the Five-Dimensional Kaluza-Klein Magnetic Monopole Spacetime}
\author{E. R. Bezerra de Mello \thanks{E-mail: emello@fisica.ufpb.br}\\
Departamento de F\'{\i}sica-CCEN\\
Universidade Federal da Para\'{\i}ba\\
58.059-970, J. Pessoa, PB\\
C. Postal 5.008\\
Brazil}
\maketitle

\begin{abstract}
In this paper we present, in a integral form, the Euclidean Green function associated with a massless scalar field in the five-dimensional Kaluza-Klein magnetic monopole superposed to a global monopole, admitting a non-trivial coupling between the field with the geometry. This Green function is expressed as the sum of two contributions: the first one related with uncharged component of the field, is similar to the Green function associated with a scalar field in a four dimensional global monopole spacetime. The second contains the information of all the other components. Using this Green function it is possible to study the vacuum polarization effects on this spacetime. Explicitly we calculate the renormalized vacuum expectation value $\langle\Phi^*(x)\Phi(x)\rangle_{Ren}$, which by its turn is also expressed as the sum of two contributions.
\end{abstract}

\newpage
\renewcommand{\thesection}{\arabic{section}.}
\section{Introduction}
A few years ago, Gross and Perry \cite{GP}, and Sorkin \cite{S} presented a soliton-like solution of the five-dimensional Kaluza-Klein theory corresponding to an Abelian magnetic monopole. As the Dirac solution, their solutions describe a gauge-dependent string singularity line, if the fifth coordinate is conveniently compactified. Moreover, their solutions depend on the parameter $m$ related with the physical magnetic charge $g$ and the the radius of the Kaluza-Klein circle $R$ by
\begin{equation}
m=g\sqrt{\pi G}=R/8 \ ,
\end{equation} 
being $G$ the Newton's constant. Also Gegenberg and Kunstatter \cite{GK} found another magnetic monopole solution. Their solutions were obtained by applying the static and Ricci-flat requirement on the filed equations. 

A global monopole is a heavy topological object formed in the phase transition of a system composed by self-coupling iso-scalar field $\phi^a$, whose original global $O(3)$ symmetry of the physical system is spontaneously broken down to $U(1)$. The scalar matter field plays the role of an order parameter which outside the monopole's core, acquires a non-vanishing value. The simplest model which gives rise to a global monopole has been proposed by Barriola and Vilenkin \cite{BV}, and is described by the Lagrangian density below
\begin{equation}
\label{GM}
{\cal{L}}=-\frac12g^{\mu\nu}\partial_\mu\phi^a\partial_\nu\phi^a-\frac14\lambda
\left(\phi^a\phi^a-\eta^2\right)^2  
\end{equation} 
with $a=1, \ 2, \ 3$ and $\eta$ being the scale energy where the symmetry is broken. The field configuration which describes a monopole is
\begin{equation}
\label{phi}
\phi^a(x)=\eta f(r)\hat{x^a} \ ,
\end{equation}
where $\hat{x^a}\hat{x^a}=1$. Coupling this matter field with the Einstein equation, a spherically symmetric regular metric tensor solution is obtained. Barriola and Vilenkin also shown that for points outside the global monopole's core the geometry of the manifold can be approximately given by the line element
\begin{equation}
\label{gm} 
ds^2=-dt^2+\frac{dr^2}{\alpha^2}+r^2(d\theta^2+\sin^2\theta d\phi^2)  
\end{equation}
with $\alpha^2=1-8\pi G\eta^2$. This line element represents a three-geometry with a solid angle deficit and a non-vanishing scalar curvature.

Recently a new solution for the Kaluza-Klein magnetic monopole in a five-dimensional global monopole spacetime has been found \cite{Mello1}. This solution corresponds to a composite topological object, i.e., an Abelian magnetic monopole superposed to a point-like global monopole. It was obtained by coupling the energy-momentum tensor associated with the five-dimensional generalization of the global monopole system, with the respective Einstein equation in the presence of an Abelian magnetic monopole. Due to the presence of the matter source, the Ricci-flat condition is no longer fulfilled. The solution can be expressed by the following line element in terms of the five-dimensional coordinates $x^A=(x^\mu, \ \Psi)=(t,\ r,\ \theta,\ \phi,\ \Psi)$:
\begin{eqnarray}
\label{S}
d{\hat{s}}^2&=&-dt^2+V(r)\left(\frac{dr^2}{\alpha^2}+r^2(d\theta^2+\sin^2\theta 
d\phi^2)\right)\nonumber\\
&+&V(r)^{-1}(d\Psi+4m(1-\cos\theta)d\phi)^2 
\end{eqnarray}
with
\begin{equation}
V(r)=1+\frac{4m}{\alpha r} \ .
\end{equation}

As has been pointed out by Gross and Perry \cite{GP}, the gauge field associated with the magnetic monopole
\begin{equation}
A_\phi=4m(1-\cos\theta) \ ,
\end{equation} 
presents a singularity at $\theta=\pi$. However, this singularity is gauge dependent if the period of the compactified coordinate $\Psi$ is equal to $16\pi m$. This is the geometric description of the Dirac quantization. Adopting this period for the extra coordinate, it is possible to provide the Wu and Yang formalism \cite{WY} to describe the four-vector potential, $A_\mu$, associated with the Abelian magnetic monopole without line of singularity. In order to do that it is necessary to construct two overlapping regions, $R_a$ and $R_b$, which cover the whole space section of the manifold. Using spherical coordinate system, with the monopole at origin the only non-vanishing components for the vector potential are
\begin{eqnarray}
\label{A}
(A_\phi)_a&=&4m(1-\cos\theta) \ ,\  R_a: \ 0 \leq \theta < \frac12\pi+\delta \ ,
\nonumber\\
(A_\phi)_b&=&-4m(1+\cos\theta) \ ,\ R_b: \ \frac12\pi-\delta< \theta \leq \pi \ , 
\end{eqnarray}
with $0<\delta <\pi/2$. In the overlapping region, $R_{ab}$, the non-vanishing 
components are related by a gauge transformation. Using the appropriate normalization
factor \cite{GP}, one can rewrite the above vector potential in terms of the 
physical one, $A^{ph}_\phi$:
\begin{equation}
\sqrt{16\pi G}(A^{ph}_\phi)_a=\sqrt{16\pi G}\left[(A^{ph}_\phi)_b+
\frac ieS\partial_\phi S^{-1}\right] \ ,
\end{equation}
where $S=e^{2iq\phi}$, $q=eg=n/2$ in units $\hbar=c=1$ and $g$ being the monopole strength. In terms of non-physical vector potential this gauge transformation corresponds to subtract the quantity $8m$, which compensates the changing in the fifth coordinate $\Psi'=\Psi+8m\phi$. 

Here, we shall consider the quantum analysis of a massless five-dimensional scalar field $\Theta(x)$ in the spacetime described by (\ref{S}). This field can be expanded in a Fourier series
\begin{equation}
\label{T}
\Theta(x)=\sum_{n=-\infty}^\infty e^{in\Psi/8m}\Theta^{(n)}(x^\mu)=\sum_{n=-\infty}^\infty e^{in\Psi/8m}e^{-iEt}R_n(r)Y_{lm}^q(\theta,\phi) \ ,
\end{equation}
being $Y_{lm}^q(\theta,\phi)$, with $q=n/2$, the monopole harmonic \cite{WY1} solution of the eingen-value equations below
\begin{eqnarray}
{\vec{L}}^2_q Y_{lm}^q=l(l+1)Y_{lm}^q  \  \quad{and} \ \ L_zY_{lm}^q=mY_{lm}^q \ , 
\end{eqnarray}
with $l=|q| \ , \ |q|+1 \ , \ |q|+2 \ , \ ...$ and $m=-l \ , \ -l+1 \ , \ ... \ , \ l$. Where
\begin{equation}
{\vec{L}}_q={\vec{r}}\times({\vec{p}}-e{\vec{A}})-q{\hat{r}} \ .
\end{equation} 

The Green function associated with the massless scalar field in this spacetime can be obtained by solving the differential equation
\begin{equation}
\label{G}
(\Box+\xi R)G(x,x')=-\delta^{(5)}(x,x')=-\frac{\delta^{(5)}(x^A-x'^A)}{\sqrt{-g^{(5)}}} \ ,
\end{equation}
$g^{(5)}$ being the determinant of the metric tensor. Here we have introduced a non-minimal coupling between the field with the geometry, $\xi R$, with $\xi$ being an arbitrary constant and $R$ the Ricci scalar curvature.

Green functions in a four-dimensional pointlike global monopole spacetime have been obtained in \cite{ML} and \cite{Mello2} to massless scalar and fermionic fields, respectively. Also the effect of temperature on these function has been considered in \cite{Mello3}. More recently the Green function associated with a charged massless scalar field in the global monopole spacetime superposed to an Abelian magnetic monopole, has been calculated in \cite{Mello4}. The Green function is important quantity to calculate vacuum polarization effects due to the presence of matter fields. Specifically for scalar field, the vacuum expectation value of the square of this field is formally given by
\begin{eqnarray}
\langle\Phi^2(x)\rangle=\lim_{x'\to x}G(x',x) \ .	
\end{eqnarray}
So the objective of this paper is to calculate the Green function associated with a massless scalar field in the five-dimensional spacetime, which takes into account the presence of the magnetic interaction between the matter field and the magnetic monopole in the unified formalism of Kaluza-Klein. Having this function the next steps is to calculate the vacuum polarization effect on this manifold, trying to understand the consequence of considering an extra compactified dimension.

This paper is organized as follows. In section $2$, we explicitly calculate the Euclidean Green function associated with this system. As we shall see, this Green function, as the scalar field, will be expressed in terms of an infinite sum of all the Fourier components associated with the quantum number $n$. For the component $n=0$, the respective Green function, $G^{(0)}(x,x')$, is similar to the scalar Green function in a four-dimensional global monopole spacetime. As to the other components, the respective Green function, $\bar{G}(x,x')$, will be expressed as the summation of all the $n\neq 0$ component $G^{(n)}(x,x')$. Although we have not found in the literature an explicit expression which provides to express $\bar{G}(x,x')$ in terms of any kind of special function, we were able to furnish an integral representation to it. In subsection $2.1$, we calculate the renormalized vacuum expectation value of the square of the scalar field, $\langle\Phi(x)^*\Phi(x)\rangle$. We show that this expectation value can be expressed as the sum of two contributions: the first, being given by the $n=0$ component of the total Green function, is associated with the uncharged component of the scalar field (\ref{T}). The second contribution takes into account all the other Fourier components. In subsection $2.2$, we present a formal expression to the renormalized vacuum expectation value of the energy-momentum tensor, $\langle T^A_B(x)\rangle$. In section $3$, we present our conclusions and the most important remarks about this system. In the Appendix we present the explicit calculation of the Green $\bar{G}(x',x)$ associated with the uncharged components of the field.

\section{Euclidean Green Function}
The Euclidean Green function can be obtained by the Schwinger-DeWitt formalism as follows:
\begin{eqnarray}
\label{G1}
G_E(x,x')=\int_0^\infty \ ds \ K(x,x';s) \ ,
\end{eqnarray}
where $K(x,x';s)$ is the heat kernel, which can be expressed in terms of the sum over the complete normalized set of eigenstates of the Klein-Gordon operator. For the massless scalar field this operator reads: 
\begin{eqnarray}
\label{KG}
(\Box+\xi R)\Theta_\sigma(x)=-\sigma^2\Theta_\sigma(x)	\ .
\end{eqnarray}
So
\begin{eqnarray}
\label{K}
K(x,x';s)=\sum_{\sigma^2}\Theta^*_\sigma(x')\Theta(x)_\sigma e^{-s\sigma^2}	\ .
\end{eqnarray}

The covariant five-dimensional d'Alembertian operator       
\begin{eqnarray}
\Box=\frac1{{\sqrt{-g^{(5)}}}}\partial_A({\sqrt{-g^{(5)}}}g^{AB}\partial_B) \ ,
\end{eqnarray}
in the spacetime defined by (\ref{S}) reads
\begin{eqnarray}
\Box&=&-\partial_t^2+\frac1{V(r)}\left\{\alpha^2\left(\partial_r^2+\frac2r\partial_r\right)-\frac{{\vec{L}}^2}{r^2}-\frac{8m(1-\cos\theta)}{r^2\sin^2\theta}\partial_\phi\partial_\Psi	 \right.\nonumber\\
&+&\left.\frac{16m^2(1-\cos\theta)^2}{r^2\sin^2\theta}\partial_\Psi^2\right\}+V(r)\partial_\Psi^2	\ .
\end{eqnarray}
On the other hand the Ricci scalar curvature is:
\begin{eqnarray}
R=\frac{2(1-\alpha^2)}{V(r)r^2}	\ .
\end{eqnarray}
Admitting to the eigenfunction $\Theta_\sigma(x)$ the form
\begin{eqnarray}
\Theta_\sigma(x)=e^{-iEt}e^{in\Psi/8m}R_l^n(r)Y_{lm}^q(\theta,\phi) 	
\end{eqnarray}
and assuming $q=n/2$ it is possible to obtain the following differential equation
\begin{eqnarray}
\left\{-E^2+\frac1{V(r)}\left[\alpha^2\left(\partial_r^2+\frac2r\partial_r\right)-\frac{({\vec{L}}^2_q-q^2)}{r^2}\right]-\frac{q^2}{16m^2}V(r)\right.\nonumber\\
+\left.\frac{2\xi(\alpha^2-1)}{V(r)r^2}\right\}R_l^n(r)Y_{lm}^q(\theta,\phi)=-\sigma^2R_l^n(r)Y_{lm}^q(\theta,\phi) \ ,
\end{eqnarray}
where we have identify
\begin{eqnarray}
{\vec{L}}^2_q={\vec{L}}^2+\frac{2iq(1-\cos\theta)}{\sin^2\theta}\partial_\phi+ q^2\frac{(i-\cos\theta)^2}{\sin^2\theta}+q^2 \ .	
\end{eqnarray}
Admitting 
\begin{eqnarray}
\sigma^2=E^2+\alpha^2p^2+\frac{q^2}{16m^2} 	
\end{eqnarray}
we find the differential equation obeyed by the unknown radial function $R(r)$:
\begin{eqnarray}
\label{R}
\frac{d^2R(r)}{dr^2}+\frac2r\frac{dR(r)}{dr}-\frac{\mu_{lq}}{r^2}R(r)-\gamma^2V(r)^2R(r)\nonumber\\
+\gamma^2V(r)R(r)+p^2V(r)R(r)=0 \ ,	
\end{eqnarray}
with 
\begin{eqnarray}
\gamma=\frac q{4m\alpha} \ \ \quad{and} \ \ \mu_{lq}=\frac{l(l+1)-q^2-2\xi(\alpha^2-1)}{\alpha^2}	\ .
\end{eqnarray}

The solution to the above differential equation regular at origin is proportional to the Whittaker function \cite{Grad} $M_{\lambda,\beta}(2ipr)$ with $\lambda$ and $\beta$ being given in terms of the parameters $m$ and $\gamma$, and also in terms of the variable $p$. Unfortunately with this function it is not possible to present the Green function in a closed form. So at this point we should make an approximation in our calculations: we shall discard the terms proportional to $m/r$ in (\ref{R}). This means that we are considering points very far from the origin\footnote{Discarding terms proportional to $m/r$ does not reduce the system to the a charged particle in the presence of an Abelian magnetic monopole \cite{Mello4}. Here the spacetime remains being five-dimensional one, and this fact is present in the calculation of the heat kernel (\ref{K}) where it is necessary to sum all the possible value for the parameter $q=n/2$.}. In \cite{GP} is given an estimative value to the radius of the Kaluza-Klein circle: $R=3.7\times 10^{-32}$cm. So we are considering points at distance greater than $3.7\times 10^{-32}$cm. Accepting this approximation (\ref{R}) becomes 
\begin{eqnarray}
\frac{d^2R(r)}{dr^2}+\frac2r\frac{dR(r)}{dr}-\frac{\mu_{lq}}{r^2}R(r)+p^2R(r)=0 \ .	
\end{eqnarray}

The solution to the above equation regular at origin is
\begin{eqnarray}
R_l^n(r)=\frac{J_{\nu_{l,q}}(pr)}{\sqrt{pr}} \ ,	
\end{eqnarray}
being $J_\nu$ the Bessel function of order
\begin{eqnarray}
\nu_l^q=\frac1\alpha\sqrt{(l+1/2)^2-q^2-2(\alpha^2-1)(\xi-1/8)} \ .	
\end{eqnarray}
So the normalized eigenstate of the Klein-Gordon operator (\ref{KG}) is
\begin{eqnarray}
\Theta_\sigma(x)=\frac1{4\pi}\sqrt{\frac{\alpha p}{2mr}}e^{-iEt}e^{in\Psi/8m} J_{\nu_l^q}(pr)Y_{lm}^q(\theta,\phi) \ .	
\end{eqnarray}
By making a Wick rotation, $t\to i\tau$, and according to (\ref{K}) the Euclidean heat kernel is given by
\begin{eqnarray}
\label{K1}
K(x,x';s)&=&\int_{-\infty}^\infty dE\sum_{n=-\infty}^\infty \int_0^\infty  dp\sum_{l,m}\Theta_\sigma(x)\Theta^*_\sigma(x')e^{-s\sigma^2}	\nonumber\\
&=&\frac1{256\pi^{5/2}m\alpha}\frac1{\sqrt{rr'}}\frac{e^{-\frac{\Delta\tau\alpha^2+r^2+r'^2}{4\alpha^2s}}}{s^{3/2}}\sum_{n=-\infty}^\infty e^{-\frac{sn^2}{64m^2}}e^{in\frac{\Psi-\Psi'}{8m}}\times\nonumber\\
&&\sum_{l=|q|}^\infty I_{\nu_l^q}\left(\frac{rr'}{2\alpha^2s}\right)\sum_{m=-l}^l Y_{lm}^q(\theta,\phi) \left(Y_{lm}^q(\theta',\phi')\right)^* \ ,
\end{eqnarray}
$I_\nu$ being the modified Bessel function. In Ref. \cite{WY2}, Wu and Yang have derived some properties of the monopole harmonics, including the generalization of spherical harmonics addition theorem. However, because we are interested to calculate the renormalized value of the Green function in the coincidence limit, a simpler expression is obtained by taking $\theta=\theta'$ and $\phi=\phi'$ in (\ref{K1}). In Appendix A of \cite{Mello4} we have shown the simplified result to the sum of product of monopole harmonics in coincidence limit in the angular variables:
\begin{eqnarray}
\sum_{m=-l}^l Y_{lm}^q(\theta,\phi) \left(Y_{lm}^q(\theta,\phi)\right)^* =\frac{(2l+1)}{4\pi} \ .	
\end{eqnarray}
Moreover we shall take the coincidence limit in the fifth coordinate $\Psi'=\Psi$ and the Euclidean temporal coordinate $\tau'=\tau$ in (\ref{K1}). Doing this we get a simpler expression to the heat kernel which depends on the quantum number $n$ in its second power.

Now according to (\ref{G1}) we get:
\begin{eqnarray}
\label{G2}
G(x,x')&=&\frac1{256\pi^{5/2}m\alpha}\frac1{\sqrt{rr'}}\sum_{l=0}^\infty(2l+1)\int_0^\infty ds \  s^{-3/2}e^{-\Omega/4\alpha^2s}I_{\nu_l^0}(\sigma/s)\nonumber\\
&+&\frac1{128\pi^{5/2}m\alpha}\frac1{\sqrt{rr'}}\sum_{n=1}^\infty\sum_{l=q}^\infty(2l+1)\int_0^\infty ds \  s^{-3/2}e^{-\Omega/4\alpha^2s}\times\nonumber\\
&&I_{\nu_l^q}(\sigma/s)e^{-sn^2/64m^2} \ ,
\end{eqnarray}
where
\begin{eqnarray}
\Omega=r^2+r'^2 \ \quad{and} \ \ \sigma=\frac{rr'}{2\alpha^2} \ .
\end{eqnarray}

As we can see the above Green function is composed by two contributions: the first one, $G^{(0)}$, corresponds to the Fourier component $n=0$ of the scalar field operator, and the second, ${\bar{G}}$ to the other components. By using \cite{Grad} it is possible to develop the integral in $s$ for the first contribution. So we get
\begin{equation}
\label{G0}
G^{(0)}(r,r')=\frac1{128\pi^3m}\frac1{rr'}\sum_{l=0}^\infty(2l+1)Q_{\nu_l^0-1/2}(\cosh u) \ ,
\end{equation}
with $\cosh u=\frac{r^2+r'^2}{2rr'}$ and $Q_\nu$ being the Legendre function, whose integral representation is
\begin{equation}
Q_{\nu-1/2}(\cosh u)=\frac1{\sqrt{2}}\int_u^\infty \ dt \frac{e^{-\nu t}}{\sqrt{\cosh t -\cosh u}} \ .
\end{equation} 

This Green function is exactly $\frac1{16\pi m}=\frac1{2\pi R}$ times the Green function obtained in a four-dimensional spacetime (see \cite{ML}). 

As to the second contribution of (\ref{G2}), we did not find in the literature the explicit expression to the integral in the variable $s$. In Appendix A we give the procedure adopted to provide a simpler and workable expression to $\bar{G}$, which will be adopted by us in the analysis of the vacuum polarization effect. The result to this contribution is:
\begin{eqnarray}
\label{Gt}
{\bar{G}}(r,r')=\sum_{n=1}^\infty \ G^{(n)}(r,r')
\end{eqnarray}
with
\begin{eqnarray}
G^{(n)}(r,r')=\frac1{128\pi^{5/2}m\alpha}\frac1{\sqrt{rr'}}\sum_{l=n/2}^\infty(2l+1) I_n
\end{eqnarray}	
being
\begin{eqnarray}
I_n=\frac1{\sqrt{\pi\sigma}}\int_u^\infty \frac{dt \ e^{-\nu_l^q t}}{\sqrt{\cosh t - \cosh u}}\cos((n/4m)\sqrt{\sigma(\cosh t - \cosh u)}) \ .	
\end{eqnarray}
Finally we can write
\begin{eqnarray}
\label{Gn}
G^{(n)}(r,r')&=&\frac1{64\sqrt{2}\pi^3m}\frac1{rr'}\sum_{l=n/2}^\infty (2l+1)\int_u^\infty \frac{dt \ e^{-\nu_l^q t}}{\sqrt{\cosh t - \cosh u}}\times\nonumber\\
&&\cos((n/4m)\sqrt{\sigma(\cosh t - \cosh u)}) \ .	
\end{eqnarray}
As the $n=0$ component of the Green function, Eq. (\ref{G0}), it is not possible to develop the integral above and  provide a closed expression to $G^{(n)}$. The best we can do is to express it in an integral representation. Moreover (\ref{Gn}) is similar to the $n=0$ component; however its coefficient is twice bigger than the coefficient multiplying the uncharged sector.  

Another point that should be mentioned is that because of the non-trivial dependence of the parameter $\nu_l^q$ with $\alpha$ and $q$, it is not possible to develop exactly the sum in $l$ in (\ref{Gn}). On the other hand for $\xi=1/8$, the geometric series in (\ref{G0}) can be performed, providing  a simpler expression to $G^{(0)}$.

Having these Green functions, it is possible to calculate the vacuum polarization effect in the gravitational background of the composite five-dimensional Kaluza-Klein magnetic and global monopoles. 

\subsection{Computation of $\langle\Phi^*(x)\Phi(x)\rangle_{Ren.}$}
Here, in this subsection, we shall discuss the calculation of the vacuum expectation value of the square of the massless scalar field operator. This quantity is formally obtained by calculating the complete Green function in the coincidence limit:
\begin{eqnarray}
\label{P2}
\langle\Phi^*(x)\Phi(x)\rangle=\lim_{x'\to x}G(x',x)=\lim_{r'\to r}[G^{(0)}(r',r)+\bar{G}(r',r)] \ .	
\end{eqnarray}
However, this procedure provides a divergent result. In order to obtain a finite and well defined result it is necessary to introduce some renormalization procedure. In this paper we shall apply the point-splitting renormalization procedure. The basic idea of this method consists to examine the singular behavior of the above limit, identify the divergent terms and subtract them off \cite{Wald}. Let us apply this method separately for each component of (\ref{P2}).

We shall first start with the uncharged component of the vacuum expectation value, 
\begin{equation}
\langle\Phi^2(x)\rangle^{(0)}=\lim_{x'\to x}G^{(0)}(x',x) \ .
\end{equation}
As we have already said $G^{(0)}$ is proportional to the Green function obtained by Mazzitelli and Lousto in \cite{ML}. So the procedure to renormalize this component is by subtracting from $G^{(0)}$ the Hadamard function below
\begin{eqnarray}
G_H(x',x)=\frac1{16\pi m}\left\{\frac1{16\pi^2} \left[\frac2{\sigma(x',x)} +\left(\xi-\frac16\right){\cal{R}}\ln\left(\frac{\mu^2\sigma(x',x)}2\right)\right]\right\}	\ ,
\end{eqnarray}
where $\mu$ is an arbitrary cutoff energy scale, ${\cal{R}}=2(1-\alpha^2)/r^2$ the scalar curvature in the four-dimensional global monopole spacetime, and $\sigma(x',x)$ one-half of the square of the geodesic distance between $x'$ and $x$. For the radial point splitting, in our approximation we have $\sigma=(r'-r)^2/2\alpha^2$. 

Because the calculation of the vacuum expectation value (VEV) of the square of the massless scalar field operator has be developed in \cite{ML} for an arbitrary value of the non-minimal coupling $\xi$, up to the first order in the parameter $\eta^2=1-\alpha^2\ll 1$, and in an exact form for $\xi=1/8$, we shall not develop here the calculation of the VEV below 
\begin{equation}
\langle\Phi^2(x)\rangle_{Ren}^{(0)}=\lim_{r'\to r}[G^{(0)}(r,r')-G_H(r,r')] \ .
\end{equation}
We shall give only the result obtained for $\xi=1/8$:
\begin{eqnarray}
\langle\Phi^2(x)\rangle&=&\frac1{512\pi^3mr^2}\int_0^\infty\frac{dt}{\sinh(t/2)}\left[\frac{\cosh(t/2\alpha)}{\sinh^2(t/2\alpha)}-\alpha^2\frac{\cosh(t/2)}{\sinh^2(t/2)}\right.\nonumber\\	
&+&\left.\frac{(\alpha^2-1)}6e^{-t/2}\right]+\frac1{1536\pi^3r^2}(1-\alpha^2)\ln(\mu r/\alpha) \ .
\end{eqnarray}

The new calculation is to obtain the VEV associated with all charged components of the Green function. In order to get some information how we should procedure, we first analyse the singular behavior of each component $G^{(n)}(r',r)$. First of all, it is necessary to develop the summation on the angular quantum number $l$ in (\ref{Gn}):
\begin{eqnarray}
S=\sum_{l=q}^\infty (2l+1)e^{-\nu_l^q t}	\ .
\end{eqnarray}
Unfortunately it is not possible to develop this summation in an exact way even for $\xi=1/8$. Here it is necessary to adopt an approximation procedure: we shall develop and expansion  of $\nu_l^q$ in powers of $\frac{q^2}{(l+1/2)^2}<1$. Moreover, developing an extra expansion in powers of parameter $\eta^2$, we get a very large expansion as shown in \cite{Mello4}. So in order to obtain a more compact expression to the summation above, which allow us to procedure the summation in the quantum number $n$ to get $\bar{G}(r',r)$, we shall adopt $\xi=1/8$. So doing this we obtain, up to the first order in $\frac{q^2}{(l+1/2)^2}$, the following expression:
\begin{eqnarray}
S=e^{-qt/\alpha}\left[\frac q{\sinh(t/2\alpha)} +\frac{\coth(t/2\alpha)}{2\sinh^2(t/2\alpha)}+\frac{q^2t}\alpha\frac1{2\sinh(t/2\alpha)}\right]	\ ,
\end{eqnarray}
with $q=n/2$. So $G^{(n)}$ becomes:
\begin{eqnarray}
\label{Gn1}
G^{(n)}(r,r')&=&\frac1{64\sqrt{2}\pi^3m}\frac1{rr'}\int_u^\infty \frac{dt \ \cos((q/2m)\sqrt{\sigma(\cosh t - \cosh u})) }{\sqrt{\cosh t - \cosh u}} \times\nonumber\\
&&e^{-qt/\alpha}\left[\frac q{\sinh(t/2\alpha)} +\frac{\coth(t/2\alpha)}{2\sinh^2(t/2\alpha)}+\frac{q^2t}\alpha\frac1{2\sinh(t/2\alpha)}\right]	\ .	
\end{eqnarray}
Analysing $G^{(n)}$ in the limit $r'\to r$, it is possible to observe that this Green function presents  effectively a four-dimensional Hadamard singular structure. So to renormalize the VEV of the square of the $n-$component of the scalar field, we should take the following Hadamard function:
\begin{eqnarray}
G_H(x',x)=\frac1{8\pi m}\left\{\frac1{16\pi^2} \left[\frac2{\sigma(x',x)} +\bar{a}_1\ln\left(\frac{\mu^2\sigma(x',x)}2\right)\right]\right\}	\ ,
\end{eqnarray}	
being $\bar{a}_1=(\xi-1/6){\cal{R}}+(2q/R)^2$, with $\xi=1/8$. ${\cal{R}}$ is the scalar curvature and $R=8m$ the radius of the circle in the fifth dimension. So, although the singular structure of the Green function be of the same type as the four-dimensional Hadamard one, the coefficient $\bar{a}_1$ in the above function presents an extra contribution proportional to the inverse of the square of the radius when compared with the expression given in the literature to this coefficient \cite{Chris} . 

After this discussion let us return to the VEV of the square of the $n$ component of the scalar field:
\begin{eqnarray}
\langle\Phi^*(x)\Phi(x)\rangle_{Ren}^{(n)}&=&\lim_{r'\to r}[G^{(n)}(r,r')-G_H(r,r')] \nonumber\\
&=&\frac1{128\pi^3mr^2}\int_0^\infty \frac{dt}{sinh(t/2)}\left\{\cos\left(\frac{qr}{2\alpha m}\sinh(t/2)\right)e^{-qt/\alpha}\times\right.\nonumber\\
&&\left[q+\frac{coth(t/2\alpha)}2+\frac{q^2t}{2\alpha}\right]\frac1{\sinh(t/2\alpha)}-\frac{\alpha^2\coth(t/2)}{\sinh^2(t/2)}\nonumber\\
&+&\left.\left(\frac{qr}{4m}\right)^2e^{-t/2}-\frac{1-\alpha^2}{12}e^{-t/2}\right\}-\frac{q^2}{1024\pi^3m^3}\ln\left(\frac{\mu r}2\right)\nonumber\\
&+&\frac{1-\alpha^2}{768\pi^3m}\frac1{r^2}\ln\left(\frac{\mu r}2\right)  \ ,
\end{eqnarray}
with $q=n/2$, being $n=\ 1 , \ 2,\ 3, \ \ ...$ . So the above result is valid only for charged Fourier components of the scalar field. 

Now let us return to the complete Green function. According to (\ref{Gt}), $\bar{G}$ is given by developing the summation on the quantum number $n$ in (\ref{Gn1}). Doing this we found:
\begin{eqnarray}
{\bar{G}}(r,r')&=&\frac1{256\sqrt{2}\pi^3m}\frac1{r'r}\int_u^\infty \frac{dt}{\sqrt{\cosh t-\cosh u}}\frac1{\sinh(t/2\alpha)}\frac1{\cosh(t/2\alpha)-\cos\gamma}\times\nonumber\\
&&\left\{\frac{\cosh(t/2\alpha)\cos\gamma-1}{\cosh(t/2\alpha)-\cos\gamma}-\cosh(t/2\alpha)(e^{-t/2\alpha}-\cos\gamma)\right.\nonumber\\
&+&\left.\frac t{8\alpha}\frac{[\cos\gamma\sinh(t/\alpha)+\cos(2\gamma)\sinh(t/2\alpha) -3\sinh(t/2\alpha)]}{(\cosh(t/2\alpha)-\cos\gamma)^2}\right\}	
\end{eqnarray}
with
\begin{eqnarray}
\gamma=\frac1{4\alpha m}\sqrt{\frac{r'r}2}\sqrt{\cosh t -\cosh u} \ \quad{and} \  \cosh u=\frac{r^2+r'^2}{2rr'} \ . 	
\end{eqnarray}

Now, at this point, we have to analyze the singular behavior of $\bar{G}$. Taking $r'\to r$, we observe that its singular behavior becomes more severe than for $G^{(n)}(r',r)$. In fact it is possible to show that in this limit we have
\begin{equation}
\label{Gb}
{\bar{G}}(r,r')\approx\frac1{\sigma^{3/2}}+\frac1{\sigma}+\frac1{\sigma^{1/2}}+\ln(\sigma) \ ,
\end{equation}
with $\sigma=(r-r')^2/2\alpha^2$. So it contains the structure of the five-dimensional Hadamard function 
\begin{equation}
G^{(5)}(r,r´)\approx \frac1{\sigma^{3/2}}+\frac1{\sigma^{1/2}} \ ,
\end{equation}
and also the four-dimensional Hadamard function
\begin{equation}
G^{(4)}(r,r')\approx \frac1{\sigma}+\ln(\sigma) \ ,
\end{equation}
at the same time. If we are inclined to provide a finite and well defined vacuum expectation value of the square of the sum of all charged components of the scalar field operator, we have to extract all the divergences of $\bar{G}$. We shall do this by subtracting from this Green function the "Hadamard" one which presents the same kind of singularity as given in (\ref{Gb}). So we have:
\begin{eqnarray}
\label{PP}
\langle\Phi^*(x)\Phi(x)\rangle_{Ren}=\lim_{r'\to r}[{\bar{G}}(r',r)-{\bar{G}}_H(r',r)] \ ,	
\end{eqnarray}
with
\begin{eqnarray}
\label{GH}
{\bar{G}}_H(r',r)=\frac{c_3}{(r'-r)^3}+\frac{c_2}{(r'-r)^2}+\frac{c_1}{(r'-r)}+\frac{c_0}{r^2}\ln\left(\frac{\mu^2(r'-r)^2}{4\alpha^2}\right)	\ .
\end{eqnarray}
The coefficients above will be determined appropriately by imposing that (\ref{PP}) be finite. Expressing all the singular terms by using the identities
\begin{eqnarray}
\frac1{(r'-r)^d}=\frac{\sqrt{2}\Gamma(\frac{d+1}d)}{2^d(r'r)^{d/2}{\sqrt{\pi}}\ \Gamma(\frac d2)}	  \int_u^\infty\frac{dt}{\sqrt{\cosh t-\cosh u}}\frac{\cosh (t/2)}{\sinh^d(t/2)} 
\end{eqnarray}
and
\begin{eqnarray}
\ln\left(\frac{\mu^2(r'-r)^2}{4\alpha^2}\right)=\ln\left(\frac{\mu^2(r'+r)^2}{4\alpha^2}\right)	-\frac2{\sqrt{2}}\int_u^\infty \frac{dt}{\sqrt{\cosh t-\cosh u}}e^{-t/2} \ ,
\end{eqnarray}
it is possible to obtain all coefficients of (\ref{GH}) according to our requirement. These coefficients are long ones; however, because of the approximation adopted in the beginning of calculation, they can be written shortly as:
\begin{eqnarray}
c_3&\approx& -\frac{m\alpha^3}{4\pi^2\sqrt{r'r}} \ , \nonumber\\
c_2&\approx& \frac{m\alpha^2}{2\pi^3r'r} \ , \nonumber\\
c_1&\approx& -\frac{7\alpha}{1536m\pi^2\sqrt{r'r}} \ , \nonumber\\
c_0&\approx& - \frac1{3072m\pi^3} \ .	
\end{eqnarray}

At this point two important remarks should be mentioned: $i)$ We did not find in the literature expressions to the coefficients of the adiabatic expansion of the Hadamard function for this five-dimensional spacetime, which presents a compactified dimension. $ii)$ Consequently we do not have any geometric explanation to them. They have been found to provide a finite result to the (\ref{PP}). The general structure to this VEV is:
\begin{eqnarray}
\langle\Phi^*(x)\Phi(x)\rangle_{Ren}=\frac A{mr^2}+\frac {Bm}{r^4}+ \frac1{1536m\pi^3}\frac1{r^2}\ln(\mu r/\alpha)	
\end{eqnarray}

\subsection{Dimensional Analysis of $\langle T_A^B\rangle$}
The energy-momentum tensor associated with scalar field in a $n$ dimensional curved spacetime is given in \cite{BD}. For this five-dimensional spacetime, considering a massless field and $\xi=1/8$ it reads:
\begin{eqnarray}
T_{AB}(x)&=&\frac34\nabla_A\Phi\nabla_B\Phi-\frac14g_{AB}g^{CD}\nabla_C\Phi \nabla_B\Phi-\frac14(\nabla_A\nabla_B\Phi)\Phi\nonumber\\
&+&\frac1{32}g_{AB}{\cal{R}}\Phi^2- \frac18{\cal{R}}_{AB}\Phi^2 \ ,	
\end{eqnarray}
where ${\cal{R}}_{AB}$ and ${\cal{R}}$ are the Ricci tensor and the scalar curvature, respectively.

The vacuum expectation value of the energy-momentum tensor operator is formally given by
\begin{eqnarray}
\langle T_{AB}(x)\rangle=\lim_{x'\to x}{\cal{D}}_{AB'}(x,x')G(x',x) \ .	
\end{eqnarray}
The non-local bivector differential operator, ${\cal{D}}_{AB'}(x',x)$, for this case reads:
\begin{eqnarray}
{\cal{D}}_{AB'}(x',x)&=&\frac34\nabla_A\nabla_{B'}-\frac14g_{AB}(x)g^{CD}(x',x)\nabla_C\nabla_{D'}\nonumber\\
&-&\frac18\left[\nabla_A\nabla_B+\nabla_{A'}\nabla_{B'}\right]+\frac1{32}g_{AB}{\cal{R}}(x)- \frac18{\cal{R}}_{AB}(x)  \ .
\end{eqnarray}
The primes denote the derivative acting at $x'$ rather than at $x$. 

The calculation of the VEV above provides a divergent result. So in order to obtain a finite and well defined result we must apply some renormalization procedure. Adopting the point-splliting renormalization one, we subtract from the Green function, $G(x',x)$, the Hadamard one, $G_H(x',x)$:
\begin{eqnarray}
\label{Te}
\langle T_{AB}(x)\rangle_{Ren}=\lim_{x'\to x}{\cal{D}}_{AB'}(x,x')[G(x',x)-G_H(x',x)] \ .	
\end{eqnarray}

This quantity must be conserved,
\begin{eqnarray}
\nabla_A\langle T_B^A(x)\rangle_{Ren}=0 \ .	
\end{eqnarray}
As to the trace anomaly, in principle it exists only for even-dimensional spacetime. In fact, as has been shown by Christensen \cite{Chris1} the trace of the renormalized VEV of the energy-momentum tensor is given by
\begin{eqnarray}
\langle T^\mu_\mu(x)\rangle_{Ren}=\frac1{(4\pi)^{n/2}}a_{n/2}(x) \ ,
\end{eqnarray}
for a even $n$ dimensional spacetime\footnote{The analysis of the vacuum polarization associated with massless scalar field in five and six-dimensional global monopole spacetime have been developed by us in \cite{Mello5}. There it is explicitly written, up to the first order in the parameter $\eta^2=1-\alpha^2$, the $a_3(x)$ coefficient, of the adiabatic expansion of the Hadamard function}. On the other hand in our previous calculations, we have found that in this five-dimensional spacetime, there appears a logarithmic term in the $\langle\Phi^*(x)\Phi(x)\rangle_{Ren}$, which should be absent in an odd-dimensional space. So we conclude that although being five-dimensional, this spacetime with the fifth coordinate compactified with period $16\pi m$, and for distance $r$ much greater than $m$, presents a four-dimensional characteristic. So it is expected a non-vanishing trace for this case, i.e., 
\begin{eqnarray}
\langle T^A_B(x)\rangle_{Ren}\neq 0 \ .
\end{eqnarray}

Another point that we want to mention is that (\ref{Te}) presents two contributions: the first one coming from the uncharged component of the scalar field, $\langle T_{AB}(x) \rangle_{Ren}^{(0)}$,  and the second coming form the other components, $\langle\bar{T}_{AB}(x) \rangle_{Ren}$. The dimensional analysis of the renormalized vacuum expectation value of the energy-momentum tensor associated with a scalar field in a four-dimensional global monopole spacetime has been developed in \cite{ML} considering this object as a pointlike one. Although the authors did not calculate explicitly this quantity, they present its general structure.

Here in this paper we also shall not calculate explicitly $\langle T_{AB}(x)\rangle_{Ren}$. However we can say that the general structure for this tensor is:
\begin{eqnarray}
\langle T_A^B(x)\rangle_{Ren}=\frac1{mr^4}\left[F_A^B(r/m,\alpha)+G_A^B(r/m,\alpha)\ln(\mu r/\alpha) \right]	\ ,
\end{eqnarray}
where $F_A^B$ and $G_A^B$ are polynomials in the ratio $r/m$ and depending on the parameter $\alpha$ which codifies the presence of the global monopole.

\section{Concluding Remarks}

In this paper we have explicitly calculate the complete Green function associated with a  massless scalar field in a five-dimensional Kaluza-Klein magnetic monopole superposed to a global monopole, for points very far from the monopole's center. This function is given in terms of two distinct contributions: one coming from the uncharged component of the scalar field, and the other from the other components:
\begin{eqnarray}
G(x',x)=G^{(0)}(x',x)+\sum_{n=1}^\infty G^{(n)}(x',x)=G^{(0)}(x',x)+{\bar{G}}(x',x) \ .	
\end{eqnarray}
Having this function, it was possible to analyse the vacuum polarization effects due to this field in this gravitational background. Explicitly we analysed the vacuum expectation value of the square of the field. This quantity is formally given by
\begin{eqnarray}
\langle\Phi^*(x)\Phi(x) \rangle=\lim_{x'\to x}G(x',x)=\lim_{x'\to x}[G^{(0)}(x',x)+ {\bar{G}}(x',x)] \ .	
\end{eqnarray}
However because this quantity provides a divergent result, to obtain a finite and well defined result, we must apply some renormalization procedure. We have applied the point-splliting renormalization procedure. As we have said before, the basic idea of this method consists to examine the singular behavior and subtract the divergent terms off, getting a finite result. We did this in a systematic way, by subtracting from the Green function the "Hadamard" one. This procedure has been developed separately for the two distinct contributions: For the first one, we observed that the singular behavior of $G^{(0)}(x',x)$ is similar to the four-dimensional Green function. However for the second contribution, the singular behavior of ${\bar{G}}(x',x)$ contains structure of the five and four-dimensional Green function. 

The Hadamard function for the first contribution could be constructed by knowing the Hadamard function for an ordinary four-dimensional spacetime, $G_H^{(4)}(x',x)$; however the second case presents a new structure. So we had to construct the respective Hadamard function in a appropriate way, ${\bar{G}}_H(x',x)$. So the renormalized vacuum expectation value of the square of the scalar field becomes:
\begin{eqnarray}
\langle\Phi^*(x)\Phi(x) \rangle_{Ren}=\lim_{x'\to x}[G^{(0)}(x',x)-G_H^{(4)}(x',x)]+
\lim_{x'\to x}[{\bar{G}}(x',x)-{\bar{G}}_H(x',x)] \ .
\end{eqnarray} 
By these analysis we may conclude that although being five-dimensional, because of the compactification of the fifth coordinates, with period $16\pi m$, and also because we are considering points at distance to the monopole, $r$, much greater than $m$, this space presents four-dimensional behavior in the singular behavior of the Green function. Consequently, there appears a logarithmic contribution in the above renormalized VEV, which presents an arbitrary energy scale $\mu$. 

We also analyse the general structure of the renormalized vacuum expectation value of the energy-momentum tensor, $\langle T_{AB}(x)\rangle_{Ren}$. By dimensional arguments and also by the result obtained in previous analysis, we infer that this quantity behaves as 
\begin{eqnarray}
\langle T_A^B(x)\rangle_{Ren}=\frac1{mr^4}\left[F_A^B(r/m,\alpha)+G_A^B(r/m,\alpha)\ln(\mu r/\alpha) \right]	\nonumber \ .
\end{eqnarray}

Unfortunately we did not calculate the explicit expressions to the two tensors above. By inspection we can see that they present dependence on the ratio $r/m$ up to second power.

{\bf{Acknowledgments}}
The author thanks to Conselho Nacional de Desenvolvimento Cient\'{\i}fico e Tecnol\'ogico (CNPq.) for partial financial support.

\section{Appendix A: Calculation of the Second Contribution of (\ref{G2})} 
The second contribution to the Green function given by (\ref{G2}), $\bar{G}(r,r')$, is given by
\begin{eqnarray}
{\bar{G}}(r,r')=\sum_{n=1}^\infty \ G^{(n)}(r,r')
\end{eqnarray}
with
\begin{eqnarray}
G^{(n)}(r,r')=\frac1{128\pi^{5/2}m\alpha}\frac1{\sqrt{rr'}}\sum_{l=q}^\infty(2l+1) I_n
\end{eqnarray}	
being
\begin{eqnarray}
I_n=\int_0^\infty ds \  s^{-3/2}e^{-\frac{\Omega}{4\alpha^2s}}I_{\nu_l^q}(\sigma/s)e^{-\frac{sn^2}{64m^2}}
\end{eqnarray}
with
\begin{eqnarray}
\Omega=r^2+r'^2 \ \quad{and} \ \ \sigma=\frac{rr'}{2\alpha^2} \ .
\end{eqnarray}

Now let us first define a new variable $y=\sigma/s$. So we obtain
\begin{eqnarray}
I_n=\frac1{\sigma^{1/2}}\int _0^\infty  \frac{dy}{\sqrt{y}}\ e^{-y\frac{\Omega}{4\alpha^2\sigma}} I_{\nu_l^q}(y)e^{-(\sigma/y)(n/8m)^2} \ .
\end{eqnarray}
Unfortunately we did not find such integral in the literature. So in order to develop an expression to the Green function we expand $e^{-\sigma n^2/64m^2y}$ in a series power getting
\begin{eqnarray}
\label{In}
I_n=\frac1{\sigma^{1/2}}\sum_{k=0}^\infty\frac{[-(n/8m)^2\sigma]^k}{k!}\int _0^\infty  dy\ y^{-k-1/2} e^{-y\frac{\Omega}{4\alpha^2\sigma}} I_{\nu_l^q}(y)	\ .
\end{eqnarray}
On page 716 of \cite{Grad}, there is a similar formula for the above integral; however in that Table there is a condition on the order of the modified Bessel function and the power of the variable $y$, that is not satisfied by the integrand of (\ref{In}), i.e., $\nu_l^q +1/2-k$ is not always a positive number. However adopting the correspondent formula, we obtain
\begin{eqnarray}
I_n=\sqrt{\frac2{\sigma\pi}}\sum_{k=0}^\infty\frac{[(n/8m)^2\sigma]^k}{k!}Q_{{\nu_l^q}-1/2}^{-k}(\cosh u)\sinh^k u\ .
\end{eqnarray}
Now substituting the integral representation below to the Legendre function \cite{Grad}
\begin{eqnarray}
Q_\nu^\mu(\cosh u)=\sqrt{\frac\pi2} \ \frac{e^{\mu\pi i}\sinh^\mu u} {\Gamma(1/2-\mu)}\int_u^\infty \frac{dt \ e^{-(\nu+1/2)t}}{(\cosh t - \cosh u)^{\mu+1/2}} \ ,	
\end{eqnarray}
we get
\begin{eqnarray}
I_n=\frac1{\sqrt{\pi\sigma}}\sum_{k=0}^\infty\frac{[-(n/4m)^2\sigma]^k}{(2k)!}\int_u^\infty	\frac{dt \ e^{-\nu_l^q t}}{\sqrt{\cosh t - \cosh u}}(\cosh t - \cosh u)^k \ .
\end{eqnarray}
Although each integral of the series diverges for $k\geq\nu_l^q+1/2$, interchanging the sum with the integral, we see that total the series obtained is the series of the cosine, so we finally get
\begin{eqnarray}
I_n=\frac1{\sqrt{\pi\sigma}}\int_u^\infty \frac{dt \ e^{-\nu_l^q t}}{\sqrt{\cosh t - \cosh u}}\cos((n/4m)\sqrt{\sigma(\cosh t - \cosh u})) \ .	
\end{eqnarray}

\end{document}